\documentclass[twocolumn,showpacs,preprintnumbers,amsmath,amssymb]{revtex4}

\usepackage{graphicx}
\usepackage{dcolumn}
\usepackage{bm}

\begin{document}

\title{On an Irreducible Theory of Complex Systems}

\author{Victor Korotkikh}
\email{v.korotkikh@cqu.edu.au}

\affiliation{Faculty of Business and Informatics\\
Central Queensland University\\
Mackay, Queensland 4740, Australia}

\author{Galina Korotkikh}

\email{g.korotkikh@cqu.edu.au}

\affiliation{Faculty of Business and Informatics\\
Central Queensland University\\
Mackay, Queensland 4740, Australia}


\begin{abstract}

In the paper we present results to develop an irreducible theory
of complex systems in terms of self-organization processes of
prime integer relations. Based on the integers and controlled by
arithmetic only the self-organization processes can describe
complex systems by information not requiring further explanations.
Important properties of the description are revealed. It points to
a special type of correlations that do not depend on the distances
between parts, local times and physical signals and thus proposes
a perspective on quantum entanglement. Through a concept of
structural complexity the description also computationally
suggests the possibility of a general optimality condition of
complex systems. The computational experiments indicate that the
performance of a complex system may behave as a concave function
of the structural complexity. A connection between the optimality
condition and the majorization principle in quantum algorithms is
identified. A global symmetry of complex systems belonging to the
system as a whole, but not necessarily applying to its embedded
parts is presented. As arithmetic fully determines the breaking of
the global symmetry, there is no further need to explain why the
resulting gauge forces exist the way they do and not even slightly
different.

\end{abstract}


\maketitle

\section{Introduction}

Complex systems profoundly change human activities of the day and
may be of strategic interest. As a result, it becomes increasingly
important to have confidence in the theory of complex systems.
Ultimately, this calls for clear explanations why the foundations
of the theory are valid in the first place. The ideal situation
would be to have an irreducible theory of complex systems not
requiring a deeper explanatory base in principle. But the question
arises: where could such a theory come from, when even the concept
of spacetime is questioned \cite{Smolin_1} as a fundamental
entity. As a possible answer it is suggested that the concept of
integers may take responsibility in the search for an irreducible
theory of complex systems \cite{Korotkikh_1}. It is shown that
complex systems can be described in terms of self-organization
processes of prime integer relations \cite{Korotkikh_1},
\cite{Korotkikh_2}. Based on the integers and controlled by
arithmetic only the self-organization processes can describe
complex systems by information not requiring further explanations.
This gives the possibility to develop an irreducible theory of
complex systems. In the paper we present results to progress in
this direction.

\section{Invariant Quantities of a Complex System and Underlying
Correlations}

To understand a complex system we consider the dynamics of the
elementary parts and focus on the correlations preserving certain
quantities of the complex system \cite{Korotkikh_1},
\cite{Korotkikh_2}.

Let $I$ be an integer alphabet and
$$
I_{N} = \{x = x_{1}...x_{N},
x_{i} \in I, i = 1,...,N \}
$$
be the set of sequences of length $N \geq 2$. We consider $N$
elementary parts $P_{i}, i = 1,...,N$ with the state of an element
$P_{i}$ specified in its reference frame by a generalized
coordinate $x_{i} \in I, i = 1,...,N$ (for example, the position
of the element $P_{i}$ in space) and the state of the elements by
a sequence $x = x_{1}...x_{N} \in I_{N}$.

For a geometric representation of the sequences we use piecewise
constant functions. Let $\varepsilon > 0$ and $\delta > 0$ be
length scales of a two-dimensional lattice. Let
$$
\rho_{m\varepsilon\delta}: x \rightarrow f
$$
be a mapping that realizes the geometric representation of a
sequence $x = x_{1}...x_{N} \in I_{N}$ by associating it with a
function $f \in W_{\varepsilon\delta}[t_{m},t_{m+N}]$, denoted $f
= \rho_{m\varepsilon\delta}(x)$, such that
$$
f(t_{m}) = x_{1}\delta, \ f(t) = x_{i}\delta, \ t \in
(t_{m+i-1},t_{m+i}], \ i = 1,...,N,
$$
$$
t_{i} = i\varepsilon, \ i = m,...,m + N
$$
and whose integrals
$f^{[k]}$ satisfy the condition
$$
f^{[k]}(t_{m}) = 0, \ k = 1,2,... \ ,
$$
where $m$ is an integer. The sequence $x = x_{1}...x_{N}$ is
called a code of the function $f$ and denoted $c(f)$.

\begin{figure}
\includegraphics[width=.40\textwidth]{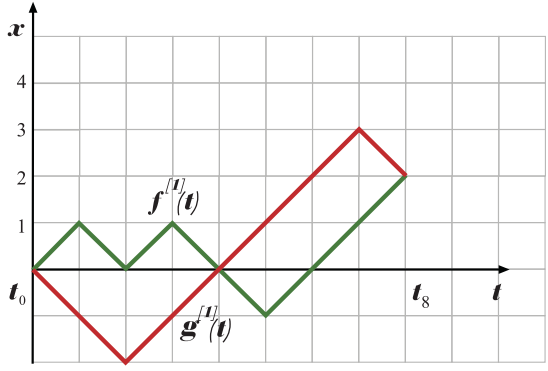}
\caption{\label{fig:one} For states $x = +1-1+1-1-1+1+1+1$ and $x'
= -1-1+1+1+1+1+1-1$ the first integrals are equal $f^{[1]}(t_{8})
= g^{[1]}(t_{8})$, where $f = \rho_{m\varepsilon\delta}(x), g =
\rho_{m\varepsilon\delta}(x')$ and $m = 0, \varepsilon = 1, \delta
= 1$. It turns out that the second integrals are also equal
$f^{[2]}(t_{8}) = g^{[2]}(t_{8})$, but the third integrals are not
$f^{[3]}(t_{8}) \neq g^{[3]}(t_{8})$. Thus, two quantities remain
invariant and $C(x,x') = 2$.}
\end{figure}

By using the geometric representation for a state $x =
x_{1}...x_{N} \in I_{N}$ of the elements $P_{i}, i = 1,..., N$ we
define quantities of the complex system as the integrals
$$
f^{[k]}(t_{m+N}) = \int_{t_{m}}^{t_{m+N}} f^{[k-1]}(t)dt, \
f^{[0]} = f, \ k = 1,2,...
$$
of a function $f^{[0]} = f = \rho_{m\varepsilon\delta}(x) \in
W_{\varepsilon\delta}[t_{m},t_{m+N}]$ \cite{Korotkikh_1}.

Remarkably, the integer code series \cite{Korotkikh_3} expresses
the definite integral
$$
f^{[k]}(t_{m+N}) = \sum_{i = 0}^{k-1}\alpha_{kmi} ((m+N)^{i}x_{1}
+ ... + (m+1)^{i}x_{N})\varepsilon^{k}\delta
$$
of a function $f \in W_{\varepsilon\delta}([t_{m},t_{m+N}])$ by
using the code $c(f) = x_{1}...x_{N}$ of the function $f$, powers
$$
(m+N)^{i},...,(m+1)^{i}, \ i = 0,...,k-1
$$
of integers $(m+N),...,(m+1)$ and combinatorial coefficients
$$
\alpha_{kmi} =
((-1)^{k-i-1}(m+1)^{k-i}+(-1)^{k-i}m^{k-i})/(k-i)!i!,
$$
where $k \geq 1$ and $i = 0,...,k-1$.

We consider the correlations conserving the quantities
\begin{equation}
\label{ST3} f^{[k]}(t_{m+N}) = g^{[k]}(t_{m+N}), \ \ \ k = 1,...,
C(x,x'),
\end{equation}
\begin{equation}
\label{ST4} f^{[C(x,x')+1]}(t_{m+N}) \neq
g^{[C(x,x')+1]}(t_{m+N}),
\end{equation}
as elements $P_{i}, i = 1,...,N$ change from one state $x =
x_{1}...x_{N} \in I_{N}$ to another $x' = x_{1}'...x_{N}' \in
I_{N}$, where $f = \rho_{m\varepsilon\delta}(x), g =
\rho_{m\varepsilon\delta}(x')$ \cite{Korotkikh_1}. The conditions
$(\ref{ST3})$ and $(\ref{ST4})$ suggest that as a complex system
changes, its parts are correlated to preserve $C(x,x')$ of the
quantities (Figure 1).

It is worthwhile to mention that the conditions $(\ref{ST3})$ and
$(\ref{ST4})$ specify a symmetry transformation possibly
determining the equation of motion of the complex system.
Therefore, once the number $C(x,x')$ of invariants is viewed as a
variable to describe an order, we may have a scheme classifying
equations of motion.

It is proved \cite{Korotkikh_1} that $C(x,x') \geq 1$ of the
quantities of a complex system remain invariant, if and only if
$C(x,x')$ equations take place
$$
(m + N)^{C(x,x')-1}\Delta x_{1} + ... + (m + 1)^{C(x,x')-1}\Delta
x_{N} = 0
$$
$$
. \qquad \qquad . \qquad  \qquad . \qquad \qquad . \qquad \qquad .
$$
$$
(m + N)^{1}\Delta x_{1} + ... + (m + 1)^{1}\Delta x_{N} = 0
$$
\begin{equation}
\label{SC1} (m + N)^{0}\Delta x_{1} + ... + (m + 1)^{0}\Delta
x_{N} = 0
\end{equation}
characterizing in view of an inequality
$$
(m + N)^{C(x,x')}\Delta x_{1} + ... + (m + 1)^{C(x,x')}\Delta
x_{N} \neq 0,
$$
the correlations between the parts of the complex system, where
$\Delta x_{i} = x_{i}' - x_{i}$ are the changes of the elements
$P_{i}, i = 1,...,N$ in their reference frames and $m$ is an
integer.

The coefficients of the system of linear equations become the
entries of the Vandermonde matrix, when the number of the
equations is $N$. This fact is important in order to prove that
$C(x,x') < N$ \cite{Korotkikh_1}.

The equations $(\ref{SC1})$ present a special type of correlations
that do not have reference to the distances between parts, local
times and physical signals. The space and non-signaling aspects of
the correlations are familiar from explanations of quantum
correlations through entanglement \cite{Gisin_1}. The time aspect
of the correlations may suggest something new into the agenda.

\section{Self-Organization Processes of Prime Integer Relations and their
Geometrization}

The equations $(\ref{SC1})$ can be also viewed as identities.
Their analysis reveals hierarchical structures of prime integer
relations in the description of complex systems
\cite{Korotkikh_1}, \cite{Korotkikh_2} (Figure 2). In the context
of the hierarchical structures it may be useful to investigate
whether the Ward identities and their generalizations
\cite{Sardanashvily_1} could be presented in a more explicit form.

The hierarchical structures underlying equations $(\ref{SC1})$ may
be in a certain superposition with each other. Namely, a prime
integer relation may simultaneously belong to a number of the
hierarchical structures. The measurements specifying the behavior
of the part controlled by the prime integer relation can propagate
through the superposition to associated prime integer relations
and thus effect other parts of the complex system.

\begin{figure}
\includegraphics[width=.50\textwidth]{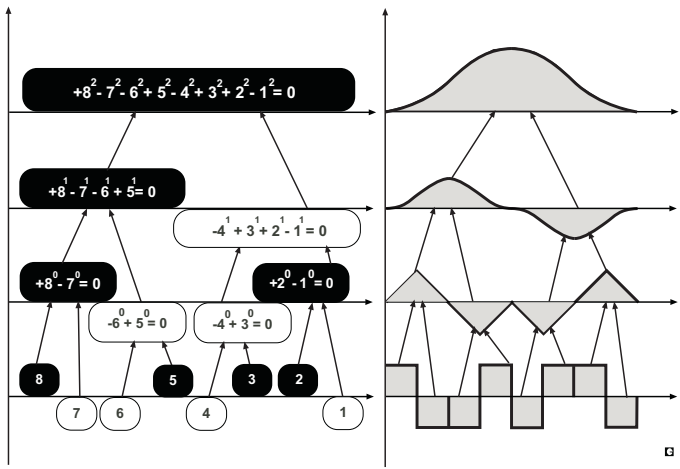}
\caption{\label{fig:two} The left side shows one of the
hierarchical structures of prime integer relations, when a complex
system has $N = 8$ elements $P_{i}, i = 1,...,8, \ x = 0 0 0 0 0 0
0 0, \ x' = +1-1-1+1-1+1+1-1, \ m = 0$ and $C(x,x') = 3$. The
hierarchical structure is built by a self-organization process of
prime integer relations and controls a correlation structure of
the complex system. The right side presents an isomorphic
hierarchical structure of geometric patterns. On scale level $0$
eight rectangles specify the dynamics of the elements $P_{i}, i =
1,...,8$. The boundary curves of the geometric patterns describe
the dynamics of the parts. All geometric patterns are symmetric
and their symmetries are interconnected. The symmetry of a
geometric pattern is global and belongs to a corresponding part as
a whole.}
\end{figure}

Through the hierarchical structures a new type of processes, i.e.,
the self-organization processes of prime integer relations, is
revealed \cite{Korotkikh_1}. Starting with integers as the
elementary building blocks and following a single principle, a
self-organization process makes up the prime integer relations of
a level of a hierarchical structure from the prime integer
relations of the lower level (Figure 2).

A prime integer relation is made as an inseparable object: if even
one of the prime integer relations is not included, then the rest
of the prime integer relations can not form an integer relation.
In other words, each and every prime integer relation involved in
the formation of a prime integer relation is critical.

The prime integer relations set limits of causality in the sense
that complex systems controlled by separate prime integer
relations have no effect on each other. Namely, if a part of a
complex system changes, then, in accordance with the prime integer
relation, the other parts change no matter how far they may be
from each other. At the same time the prime integer relation has
no causal power to influence parts of complex systems controlled
by separate prime integer relations.

An analogy between the prime integer relation and the light cone
may be suggested. While in spacetime events can be only observed
in light cones, in our approach events can only happen through the
hierarchical structure of the prime integer relations.

By using the integer code series \cite{Korotkikh_3} the prime
integer relations can be geometrized as two-dimensional geometric
patterns and the self-organization processes can be isomorphically
expressed by transformations of the patterns \cite{Korotkikh_1}.
As it becomes possible to measure a prime integer relation by an
isomorphic geometric pattern, quantities of the prime integer
relation and a complex system it describes can be defined by
quantities of the geometric pattern such as the area and the
length of its boundary curve (Figure 2).

Due to the isomorphism, the structure and the dynamics of a
complex system are combined. As self-organization processes of
prime integer relations determine the correlation structure of a
complex system, transformations of corresponding geometric
patterns may characterize its dynamics in a strong scale covariant
form \cite{Korotkikh_1}, \cite{Korotkikh_2}.

\section{Structural Complexity in Optimality Condition of
Complex Systems and Optimal Quantum Algorithms}

Despite different origin complex systems have much in common and
are investigated to satisfy universal laws. Our description points
out that the universal laws may originate not from forces in
spacetime, but through arithmetic.

There are many notions of complexity introduced in the search to
communicate the universal laws into theory and practice. The
concept of structural complexity is defined to measure the
complexity of a system in terms of self-organization processes of
prime integer relations \cite{Korotkikh_1}. In particular, as
self-organization processes of prime integer relations progress
from a level to the higher level, the system becomes more complex,
because its parts at the level are combined to make up more
complex parts at the higher level. Therefore, the higher the level
self-organization processes progress to, the greater is the
structural complexity of a corresponding complex system.

Existing concepts of complexity do not explain in general how the
performance of a complex system may depend on its complexity. To
address the situation we conducted computational experiments to
investigate whether the concept of structural complexity could
make a difference \cite{Korotkikh_4}, \cite{Korotkikh_5}.

A special optimization algorithm, as a complex system, was
developed to minimize the average distance in the travelling
salesman problem. Remarkably, for each problem the performance of
the algorithm was concave. As a result, the algorithm and a
problem were characterized by a single performance optimum. The
analysis of the performance optimums for all problems tested
revealed a relationship between the structural complexity of the
algorithm and the structural complexity of the problem
approximating it well enough by a linear function
\cite{Korotkikh_5}.

The results of the computational experiments suggest the
possibility of a general optimality condition of complex systems:

{\it A complex system demonstrates the optimal performance for a
problem, when the structural complexity of the system is in a
certain relationship with the structural complexity of the
problem.}

The optimality condition presents the structural complexity of a
system as a key to its optimization. Indeed, according to the
optimality condition the optimal result can be obtained as long as
the structural complexity of the system is properly related with
the structural complexity of the problem. From this perspective
the optimization of a system should be primarily concerned with
the control of the structural complexity of the system to match
the structural complexity of the environment.

The computational results also indicate that the performance of a
complex system may behave as a concave function of the structural
complexity. Once the structural complexity would be controlled as
a single entity, the optimization of a complex system could be
potentially reduced to a one-dimensional concave optimization
irrespective of the number of variables involved its description.

In the search to identify a mathematical structure underlying
optimal quantum algorithms the majorization principle emerges as a
necessary condition for efficiency in quantum computational
processes \cite{Latorre_1}. We find a connection between the
optimality condition and the majorization principle in quantum
algorithms.

According to the majorization principle in an optimal quantum
algorithm the probability distribution associated to the quantum
state has to be step-by-step majorized until it is maximally
ordered. This means that an optimal quantum algorithm works in
such a way that the probability distribution $p_{k+1}$ at step
$k+1$ majorizes $p_{k} \prec p_{k+1}$ the probability distribution
$p_{k}$ at step $k$. There are special conditions in place for the
probability distribution $p_{k+1}$ to majorize the probability
distribution $p_{k}$ with intuitive meaning that the distribution
$p_{k}$ is more disordered than $p_{k+1}$ \cite{Latorre_1}.

In our description the algorithm revealing the optimality
condition also uses a similar principle, but based on the
structural complexity. The algorithm tries to work in such a way
that the structural complexity ${\bf C}_{k+1}$ of the algorithm at
step $k+1$ majorizes ${\bf C}_{k} \prec {\bf C}_{k+1}$ its
structural complexity ${\bf C}_{k+1}$ at step $k$. The concavity
of the algorithm's performance suggests efficient means to find
optimal solutions \cite{Korotkikh_5}.

\section{Global Symmetry of Complex Systems and Gauge Forces}

Our description presents a global symmetry of complex systems
through the geometric patterns of prime integer relations and
their transformations. It belongs to the system as a whole, but
does not necessarily apply to its embedded parts. The differences
between the behaviors of the parts may be interpreted through the
existence of gauge forces acting in their reference frames. As
arithmetic fully determines the breaking of the global symmetry,
there is no further need to explain why the resulting gauge forces
exist the way they do and not even slightly different.

Let us illustrate the results by a special self-organization
process of prime integer relations \cite{Korotkikh_1},
\cite{Korotkikh_2}. The left side of Figure 2 shows a hierarchical
structure of prime integer relations built by the process. It
controls a correlation structure of a complex system with states
of $N = 8$ elements $P_{i}, i = 1,...,8$ given by sequences
$$ x =
0 0 0 0 0 0 0 0, \ x' = +1 -1 -1 +1 -1 +1 +1 -1
$$
and $m = 0$. The sequence $x'$ is the initial segment of length
$8$ of the Prouhet-Thue-Morse (PTM) sequence starting with $+1$.
There is an ensemble of self-organization processes and thus
correlation structures forming the correlation structure of the
complex system associated with the states $x$ and $x'$. The
self-organization process we consider is only one of them.

\begin{figure}
\includegraphics[width=.47\textwidth]{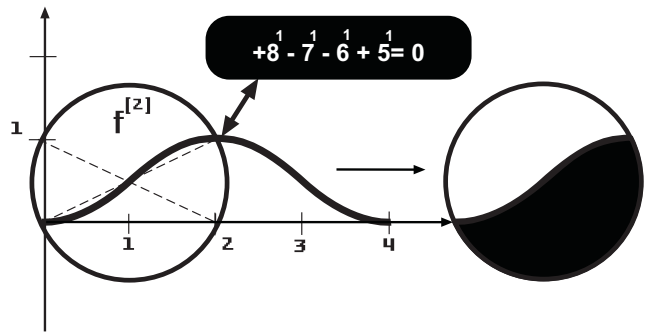}
\caption{\label{fig:three} The geometric pattern of the part
$(P_{1} \leftrightarrow P_{2}) \leftrightarrow (P_{3}
\leftrightarrow P_{4})$. From above the pattern is limited by the
boundary curve, i.e., the graph of the second integral
$f^{[2]}(t), \ t_{0} \leq t \leq t_{4}$ of the function $f$
defined on scale level $0$ (Figure 2), where $t_{i} =
i\varepsilon, \ i = 1,..., 4, \varepsilon = 1$, and it is
restricted by the $t$ axis from below. The geometric pattern is
isomorphic to the prime integer relation $+8^{1} - 7^{1} - 6^{1}+
5^{1} = 0$ and determines the dynamics. If the part deviates from
this dynamics even slightly, then some of the correlation links
provided by the prime integer relation disappear and the part
decays. The boundary curve has a special property ensuring that
the area of the geometric pattern is given as the area of a
triangle: $S = \frac{HL}{2}$, where $H$ and $L$ are the height and
the length of the geometric pattern. In the figure $H = 1$ and $L
= 4$, thus $S = 2$. The property is illustrated in yin-yang
motifs.}
\end{figure}

The right side of Figure 2 presents an isomorphic hierarchical
structure of geometric patterns. The boundary curves of the
geometric patterns determine the dynamics of parts of the complex
system. Quantities of a geometric pattern, such as its area and
the length of the boundary curve, define quantities of a
corresponding part of the complex system. The quantities of the
parts are interconnected through the transformations of the
geometric patterns.

The area of the geometric pattern has a remarkable property.
Although a geometric pattern on scale level ${\cal N}
> 1$ is bounded by an intricate curve, its area $S$, nevertheless,
can be expressed as the area of a triangle by
$$
S = \frac{HL}{2},
$$
where $H$ and $L$ are the height and the length of the geometric
pattern (Figures 2 and 3). The area has a strong scale covariant
description: the law of geometric pattern area and corresponding
quantity $E$ is the same and in the simplest possible form for all
scale levels \cite{Korotkikh_2}.

Importantly, the quantity of a complex system specified by the
height $H$ of the geometric pattern and denoted by $M$ is
determined by the area quantities $E_{l}$ and $E_{r}$ of two
parts, i.e., left and right, the complex system is made of (Figure
2). Namely, the height $H$ of the geometric pattern equals one
half of the sum of the areas of the geometric patterns
characterizing the parts forming the complex system from the lower
scale level.

Thus, we have
$$
M = \frac{E_{l} + E_{r}}{2}
$$
suggesting that, as the parts on scale level ${\cal N}-1$ form the
complex system on scale level ${\cal N}$, their $E$ quantities are
converted into the $M$ quantity of the complex system. The $M$
quantity is proportional and as such can be seen equivalent to the
$E$ quantity of the complex system
$$
E = \frac{ML}{2}.
$$

To investigate how the length $\vartheta$ of the boundary curve of
a geometric pattern could be expressed in terms of its length $L$
we try for $\varepsilon = 1$ and $\delta = 1$ a formula
\begin{equation}
\label{ADD}
\vartheta = \frac{{\cal F}L}{4},
\end{equation}
where
$$
{\cal F} = 4.6692 . . .
$$
is the Feigenbaum constant. It turns out that in the case of the
self-organization process the formula $(\ref{ADD})$ works better
and better as we proceed from scale level $1$ to scale level $3$.
This can be demonstrated by comparing the length calculated
numerically \cite{Korotkikh_2} and the same length but given by
the formula.

For the geometric pattern of the part $P_{1} \leftrightarrow
P_{2}$ at scale level $1$ the value of $\vartheta$ when obtained
computationally is $2.8284 . . . \ $. The formula $(\ref{ADD})$
for ${\cal F} = 4.6692$ and $L = 2$ gives
$$
\vartheta_{1}' = \frac{4.6692 \times 2}{4} = 2.3346.
$$
We can calculate the error as
$$
e_{1} = \frac{\mid \vartheta_{1} -
\vartheta_{1}'\mid}{\vartheta_{1}}
$$
$$
= \frac{\mid 2.8284 - 2.3346\mid}{2.8284} \approx 0.2.
$$
For the geometric pattern of the part
$$
(P_{1} \leftrightarrow P_{2}) \leftrightarrow (P_{3}
\leftrightarrow P_{4})
$$
at scale level $2$ we get
$$
\vartheta_{2} = 4.5911 . . .
$$
and by the formula $(\ref{ADD})$ for $L = 4$
$$
\vartheta_{2}' = \frac{4.6692 \times 4}{4} = 4.6692,
$$
so the error
$$
e_{2} = \frac{\mid \vartheta_{2} -
\vartheta_{2}'\mid}{\vartheta_{2}}
$$
$$
= \frac{\mid 4.5911 - 4.6692 \mid}{4.5911} \approx 0.02.
$$

Similarly, for the geometric pattern of the part
$$
((P_{1} \leftrightarrow P_{2}) \leftrightarrow (P_{3}
\leftrightarrow P_{4})) \leftrightarrow ((P_{5} \leftrightarrow
P_{6}) \leftrightarrow (P_{7} \leftrightarrow P_{8}))
$$
at scale level $3$ we have
$$
\vartheta_{3} = 9.3248. . .  \ ,
$$
$$
\vartheta_{3}' = \frac{4.6692 \times 8}{4} = 9.3384
$$
and
$$
e_{3} = \frac{\mid \vartheta_{3} -
\vartheta_{3}'\mid}{\vartheta_{3}}
$$
$$
= \frac{\mid 9.3248 -  9.3384 \mid}{9.3248} \approx 0.0015.
$$

Therefore, the formula $(\ref{ADD})$ gives roughly a ten times
better approximation to the length of the boundary curve as a
transition is made from a scale level ${\cal N}$ to the scale
level ${\cal N}+1$, where ${\cal N} = 1,2$ (Figure 2).

Starting with the elements at scale level $0$, the parts of the
correlation structure are built scale level by scale level and
thus a part of the complex system becomes a complex system itself.
From Figure 2 the symmetries of the geometric patterns and their
interconnection generated through integrations of the function can
be seen.

We consider whether the description of the dynamics of parts of a
scale level is invariant as through the formation they become
embedded in a part of the higher scale level. At scale level $2$
the second integral
$$
f^{[2]}(t), \ t_{0} \leq t \leq t_{4}, \ t_{i} = i\varepsilon, \ i
= 1,..., 4, \ \varepsilon = 1
$$
characterizes the dynamics of the part
$$
(P_{1} \leftrightarrow P_{2}) \leftrightarrow (P_{3}
\leftrightarrow P_{4}).
$$
This composite part is made up of elements $ P_{1}, P_{2}, P_{3},
P_{4}$ and parts $P_{1} \leftrightarrow P_{2}, P_{3}
\leftrightarrow P_{4}$ embedded by the formations within its
correlation structure (Figures 2 and 3).

The description of the dynamics of elements $P_{1}, P_{2}, P_{3},
P_{4}$ and  parts $P_{1} \leftrightarrow P_{2}, P_{3}
\leftrightarrow P_{4}$ within the part
$$
(P_{1} \leftrightarrow P_{2}) \leftrightarrow (P_{3}
\leftrightarrow P_{4})
$$
is invariant relative to their reference frames. In particular,
the dynamics of elements $P_{1}$ and $P_{2}$ in a reference frame
of the element $P_{1}$ is specified by
$$
f^{[2]}(t) = f_{P_{1}}^{[2]}(t_{P_{1}}) =
\frac{t_{P_{1}}^{2}}{2!},
$$
\begin{equation}
\label{X2}
t_{0} = t_{0,P_{1}} \leq t_{P_{1}} \leq t_{1,P_{1}} =
t_{1},
\end{equation}
$$
f^{[2]}(t) = f_{P_{1}}^{[2]}(t_{P_{1}}) =
-\frac{t_{P_{1}}^{2}}{2!} + 2t_{P_{1}} - 1,
$$
$$
t_{1} = t_{1,P_{1}} \leq t_{P_{1}} \leq t_{2,P_{1}} = t_{2}.
$$

The transition from the coordinate system of the element $P_{1}$
to a coordinate system of the element $P_{2}$ given by the
transformation
$$
t_{P_{2}} = - t_{P_{1}} - 2, \ \ \ f^{[2]}_{P_{2}}
= -f^{[2]}_{P_{1}} - 1
$$
shows that the characterization
\begin{equation}
\label{X3} f_{P_{2}}^{[2]}(t_{P_{2}}) = \frac{t_{P_{2}}^{2}}{2!},
\ \ \  t_{0,P_{2}} \leq t_{P_{2}} \leq t_{1,P_{2}}
\end{equation}
of the dynamics of the element $P_{2}$ is invariant, if we compare
$(\ref{X2})$ and $(\ref{X3})$.

The description is also invariant, when we consider the dynamics
of the elements $P_{3}$ and $P_{4}$ in their reference frames. The
dynamics of the element $P_{3}$ in the coordinate system of the
element $P_{1}$ is specified by
$$
f_{P_{1}}^{[2]}(t_{P_{1}}) = -\frac{t_{P_{1}}^{2}}{2!} +
2t_{P_{1}} -1,
$$
$$
t_{2,P_{1}}  \leq t_{P_{1}} \leq  t_{3,P_{1}}.
$$
The description of the dynamics of the element $P_{3}$ takes the
same form
\begin{equation}
\label{X4} f_{P_{3}}^{[2]}(t_{P_{3}}) = \frac{t_{P_{3}}^{2}}{2!},
\ \ \ t_{0,P_{3}} \leq t_{P_{3}} \leq t_{1,P_{3}}
\end{equation}
as $(\ref{X2})$ and $(\ref{X3})$, when using the transformation
$$
t_{P_{3}} = t_{P_{1}} - 2, \ \ \  f^{[2]}_{P_{3}} =
-f^{[2]}_{P_{1}} + 1,
$$
a transition to a coordinate system of the element $P_{3}$ is
made.

Similarly, the dynamics of the element $P_{4}$ in the coordinate
system of the element $P_{1}$ is specified by
$$
f_{P_{1}}^{[2]}(t_{P_{1}}) = \frac{t_{P_{1}}^{2}}{2!} - 4t_{P_{1}}
+ 8,
$$
$$
t_{3,P_{1}} \leq t_{P_{1}} \leq t_{4,P_{1}}.
$$
The transformation
$$
t_{P_{4}} = - t_{P_{1}} + 4, \ \ \  f^{[2]}_{P_{4}} =
f^{[2]}_{P_{1}}
$$
leads to a coordinate system of the element $P_{4}$ to demonstrate
that the description of the dynamics of the element $P_{4}$ has
the same form
$$
f_{P_{4}}^{[2]}(t_{P_{4}}) = \frac{t_{P_{4}}^{2}}{2!}, \ \ \
t_{0,P_{4}} \leq t_{P_{4}} \leq t_{1,P_{4}},
$$
as $(\ref{X2})$, $(\ref{X3})$ and $(\ref{X4})$.

Furthermore, descriptions of the dynamics of the parts $P_{1}
\leftrightarrow P_{2}$ and $P_{3} \leftrightarrow P_{4}$ are the
same relative to their coordinate systems. Namely, for the
dynamics of the part $P_{1} \leftrightarrow P_{2}$ in its
reference frame we have
\begin{equation}
\label{X5} f_{P_{1} \leftrightarrow P_{2}}^{[2]}(t_{P_{1}
\leftrightarrow P_{2}}) = \left\{
\begin{array}{l}
\frac{t_{P_{1} \leftrightarrow P_{2}}^{2}}{2!},
\\
\ t_{0,P_{1} \leftrightarrow P_{2}} \leq t_{P_{1} \leftrightarrow
P_{2}} \leq t_{1,P_{1} \leftrightarrow P_{2}}
\\ \\
-\frac{t_{P_{1} \leftrightarrow P_{2}}^{2}}{2!} + 2t_{P_{1}
\leftrightarrow P_{2}} - 1,
\\
\ t_{1,P_{1} \leftrightarrow P_{2}} \leq t_{P_{1} \leftrightarrow
P_{2}} \leq t_{2,P_{1} \leftrightarrow P_{2}}
\end{array}
\right.
\end{equation}

For the dynamics of the part $P_{3} \leftrightarrow P_{4}$ in the
reference frame of the part $(P_{1} \leftrightarrow P_{2})$ we
have
\begin{equation}
\label{X6}
 f_{P_{1} \leftrightarrow P_{2}}^{[2]}(t_{P_{1}
\leftrightarrow P_{2}}) = \left\{
\begin{array}{l}
-\frac{t_{P_{1} \leftrightarrow P_{2}}^{2}}{2!} + 2t_{P_{1}
\leftrightarrow P_{2}} - 1,
\\
t_{2,P_{1} \leftrightarrow P_{2}} \leq t_{P_{1} \leftrightarrow
P_{2}} \leq t_{3,P_{1} \leftrightarrow P_{2}}
\\ \\
\frac{t_{P_{1} \leftrightarrow P_{2}}^{2}}{2!} - 4t_{P_{1}
\leftrightarrow P_{2}} + 8,
\\
t_{3,P_{1} \leftrightarrow P_{2}} \leq t_{P_{1} \leftrightarrow
P_{2}} \leq t_{4,P_{1} \leftrightarrow P_{2}}
\end{array}
\right.
\end{equation}

The description $(\ref{X6})$ takes the same form
$$
f_{P_{3} \leftrightarrow P_{4}}^{[2]}(t_{P_{3} \leftrightarrow
P_{4}}) = \left\{
\begin{array}{l}
\frac{t_{P_{3} \leftrightarrow P_{4}}^{2}}{2!},
\\
t_{0,P_{3} \leftrightarrow P_{4}} \leq t_{P_{3} \leftrightarrow
P_{4}} \leq t_{1,P_{3} \leftrightarrow P_{4}}
\\ \\
-\frac{t_{P_{3} \leftrightarrow P_{4}}^{2}}{2!} + 2t_{P_{3}
\leftrightarrow P_{4}} - 1,
\\
\ t_{1,P_{3} \leftrightarrow P_{4}} \leq t_{P_{3} \leftrightarrow
P_{4}} \leq t_{2,P_{3} \leftrightarrow P_{4}}
\end{array}
\right.
$$
as $(\ref{X5})$, if under the transformation
$$
t_{P_{3} \leftrightarrow P_{4}} = t_{P_{1} \leftrightarrow P_{2}}
+ 2, \ \ \ f^{[2]}_{P_{3} \leftrightarrow P_{4}} = -f^{[2]}_{P_{1}
\leftrightarrow P_{2}} + 1,
$$
we make a transition from the reference frame of the part $P_{1}
\leftrightarrow P_{2}$ to a reference frame of the part $P_{3}
\leftrightarrow P_{4}$.

Thus, as the perspective is changed from the reference frame of
the part $P_{1} \leftrightarrow P_{2}$ to the reference frame of
the part $P_{3} \leftrightarrow P_{4}$ the description of the
dynamics remains invariant.

However, at scale level $3$ the description of the dynamics is not
invariant. In particular, the dynamics of elements $P_{1}$ and
$P_{2}$ within the part
$$
((P_{1} \leftrightarrow P_{2}) \leftrightarrow (P_{3}
\leftrightarrow P_{4})) \leftrightarrow ((P_{5} \leftrightarrow
P_{6}) \leftrightarrow (P_{7} \leftrightarrow P_{8}))
$$
relative to a coordinate system of the element $P_{1}$ can be
specified accordingly by (Figure 2)
\begin{equation}
\label{X7} f_{P_{1}}^{[3]}(t_{P_{1}}) = \frac{t_{P_{1}}^{3}}{3!},
\ \ \ t_{0,P_{1}} \leq t \leq t_{1,P_{1}},
\end{equation}
$$
f_{P_{1}}^{[3]}(t_{P_{1}}) = -\frac{t_{P_{1}}^{3}}{3!} +
t_{P_{1}}^{2} - t_{P_{1}} + \frac{1}{3}, \  t_{1,P_{1}} \leq
t_{P_{1}} \leq t_{2,P_{1}}.
$$

The transitions from the coordinate systems of the element $P_{1}$
to the coordinate systems of the element $P_{2}$ do not preserve
the form $(\ref{X7})$. For example, if under the transformation
$$
t_{P_{2}} = t_{P_{1}} + 2, \ f^{[3]}_{P_{2}} = -f^{[3]}_{P_{1}} +
1
$$
the perspective is changed from the coordinate system of the
element $P_{1}$ to a coordinate system of the element $P_{2}$,
then it turns out that the description of the dynamics
$(\ref{X7})$ is not invariant
$$
f^{[2]}(t) = f_{P_{2}}^{[3]}(t_{P_{2}}) = \frac{t_{P_{2}}^{3}}{3!}
- t_{P_{2}}, \ \ \ t_{1,P_{2}} \leq t_{P_{1}} \leq t_{2,P_{2}}
$$
due to the additional linear term $-t_{P_{2}}$.

Therefore, on scale level 3 arithmetic determines the different
dynamics of the elements $P_{1}$ and $P_{2}$. Information about
the difference might be obtained from two observers positioned at
the coordinate system of the element $P_{1}$ and the coordinate
system of the element $P_{2}$ respectively. As one observer would
report about the dynamics of the element $P_{1}$ and the other
about the dynamics of the element $P_{2}$, we could find the
difference and interpret it through the existence of a gauge force
$F$ acting on the element $P_{2}$ in its coordinate system to the
effect of the linear term $\chi(F) = - t_{P_{2}}$
$$
f_{P_{2}}^{[3]}(t_{P_{2}}) = \frac{t_{P_{2}}^{3}}{3!} - \chi(F),
$$
$$
t_{0,P_{2}} \leq t_{P_{2}} \leq t_{1,P_{2}}.
$$

The introduction of the gauge force $F$ restores the local
symmetry
\begin{equation}
\label{X10} f_{P_{2}}^{[3]}(t_{P_{2}}) = \frac{t_{P_{2}}^{3}}{3!},
\ \ \ t_{0,P_{2}} \leq t_{P_{2}} \leq t_{1,P_{2}}
\end{equation}
as we can see comparing $(\ref{X7})$ and $(\ref{X10})$.

The results can be schematically expressed as follows
\medskip

\centerline{\it Arithmetic $\rightarrow$}
\smallskip

\centerline{\it Prime integer relations in control}
\smallskip

\centerline{\it of correlation structures of complex systems
$\leftrightarrow$}
\smallskip

\centerline{\it Global symmetry: geometric patterns in control}
\smallskip

\centerline{\it of the dynamics of complex systems $\rightarrow$}
\smallskip

\centerline{\it $\rightarrow$ Not locally invariant descriptions}
\smallskip

\centerline{\it of embedded parts of complex systems
$\leftrightarrow$}
\smallskip

\centerline{\it $\leftrightarrow$ Gauge forces to restore local
symmetries}
\medskip

To discuss how the gauge forces resulting from the breaking of the
global symmetry could be quantitatively classified, we consider
whether the scale levels in our description could be subdivided
into groups of successive scale levels so that the description of
complex systems would be group scale invariant. If it would be the
case then the classification of the gauge forces could be made
possible by focusing on any group of scale levels. Every other
group would have the same classification although expressed in its
own terms.

Remarkably, as far as the self-organization process of prime
integer relations is concerned, it turns out that the scale levels
can be subdivided into groups of {\it three} successive levels,
where the description of complex systems remains the same. In
particular, using the renormalization group transformation applied
to $\varepsilon, \delta$
$$ \varepsilon' =
2^{3}\varepsilon, \ \ \ \delta' = \varepsilon^{3}\delta,
$$
and elements $P_{1}, P_{2}, . . . $
$$
P_{1}' = ((P_{1} \leftrightarrow P_{2}) \leftrightarrow (P_{3}
\leftrightarrow P_{4}))
$$
$$
\leftrightarrow ((P_{5} \leftrightarrow P_{6}) \leftrightarrow
(P_{7} \leftrightarrow P_{8})),
$$
$$
P_{2}' = ((P_{9} \leftrightarrow P_{10}) \leftrightarrow (P_{11}
\leftrightarrow P_{12}))
$$
$$
\leftrightarrow ((P_{13} \leftrightarrow P_{14}) \leftrightarrow
(P_{15} \leftrightarrow P_{16})),
$$
\centerline{. \ \ \ . \ \ \ . \ \ \ . \ \ \ . \ \ \ . }
\centerline{ }
we can obtain the same description of complex
systems on scale levels $4,5,6$ as for complex systems on scale
levels $1,2,3$. The difference is that the description on scale
levels $4,5,6$ is given in terms of $\varepsilon'$ and $\delta'$
with parts $P_{1}', P_{2}',... \ $ as the elements, while on scale
levels $1,2,3$ it is given in terms of $\varepsilon$, $\delta$ and
elements $P_{1}, P_{2},... \ $. The situation repeats for scale
levels $7,8,9$ and so on \cite{Korotkikh_2}.

\section{Conclusions}

In the paper we have presented results to develop an irreducible
theory of complex systems in terms of self-organization processes
of prime integer relations. Based on the integers and controlled
by arithmetic only the self-organization processes can describe
complex systems by information not requiring further explanations.
The following properties have been revealed.

First, the description points to a special type of correlations
that do not depend on the distances between parts, local times and
physical signals. Apart from the time aspect such correlations are
known in quantum physics and attributed to quantum entanglement
\cite{Gisin_1}. Thus, the description proposes a perspective on
quantum entanglement suggesting to include the time aspect into
the agenda.

Second, through a concept of structural complexity the description
computationally reveals the possibility of a general optimality
condition of complex systems. According to the optimality
condition the optimal result can be obtained as long as the
structural complexity of the system is properly related with the
structural complexity of the problem. The experiments also
indicate that the performance of a complex system may behave as a
concave function of the structural complexity. Therefore, once the
structural complexity would be controlled as a single entity, the
optimization of a complex system could be potentially reduced to a
one-dimensional concave optimization irrespective of the number of
variables involved in its description.

In the search to identify a mathematical structure underlying
optimal quantum algorithms the majorization principle emerges as a
necessary condition for efficiency in quantum computational
processes \cite{Latorre_1}. A connection between the majorization
principle in quantum algorithms and the optimality condition has
been identified. While the quantum majorization principle suggests
that the computational process should stop when the probability
distribution is maximally ordered, it does not however specify
what this order actually means in the context of a particular
problem. At the same time our approach is clear on this matter: to
obtain the performance maximum the computational process should
stop when its structural complexity is in a certain relationship
with the structural complexity of the problem.

Third, the description introduces a global symmetry of complex
systems that belongs to the system as a whole, but does not
necessarily apply to its embedded parts. The breaking of the
global symmetry may be interpreted through the existence of gauge
forces. There is no further need to explain why the resulting
gauge forces exist the way they do and not even slightly different
as they are fully determined by arithmetic.

\bibliography{apssamp}

\end{document}